\documentclass[prd,twocolumn,groupedaddress,showpacs,nofootinbib]{revtex4}
\usepackage{graphicx}
\usepackage{dcolumn}
\usepackage{bm}
\usepackage{amssymb}
\usepackage{mathrsfs}
\usepackage{amsmath}
\usepackage{epsfig}
\usepackage[dvips]{color}
\usepackage{hhline}

\begin{document}

\title{Leptogenesis with Linear, Inverse or Double Seesaw}

\author{Pei-Hong Gu$^{1}_{}$}
\email{peihong.gu@mpi-hd.mpg.de}

\author{Utpal Sarkar$^{2}_{}$}
\email{utpal@prl.res.in}

\affiliation{$^{1}_{}$Max-Planck-Institut f\"{u}r Kernphysik,
Saupfercheckweg
1, 69117 Heidelberg, Germany\\
$^{2}_{}$Physical Research Laboratory, Ahmedabad 380009, India}

\begin{abstract}

The left-right symmetric model with doublet and bi-doublet Higgs
scalars can accommodate linear, inverse or double seesaw for
generating small neutrino masses in the presence of three singlet
fermions. If the singlet fermions have small Majorana masses, they
can form three pairs of quasi-degenerate Majorana fermions with
three right-handed neutrinos. The decays of the quasi-degenerate
Majorana fermions can realize the resonant leptogenesis.
Alternatively, the right-handed neutrinos can obtain seesaw
suppressed Majorana masses if the singlet fermions are very heavy.
In this case leptogenesis, with or without resonant effect, is
allowed in the decays of the right-handed neutrinos.

\end{abstract}


\maketitle

\section{Introduction}

Neutrinos are massless in the standard model (SM). Under the SM
gauge symmetry, one can introduce right-handed neutrinos
\cite{minkowski1977,yanadida1979,gps1979,glashow1980,ms1980} or
Higgs triplet(s) \cite{mw1980} to accommodate the seesaw
\cite{minkowski1977,yanadida1979,gps1979,glashow1980,ms1980}
mechanism for generating the small neutrino masses naturally. A more
attractive scheme is to consider the left-right symmetric extension
of the SM. The left-right symmetric models \cite{ps1974}, based on the
gauge groups $SU(2)_{L}^{}\times SU(2)_{R}^{}\times U(1)_{B-L}^{}$
have a number of attractive features, such as the natural
explanation of weak hypercharge in terms of lepton and baryon
numbers, the origin of parity violation, the existence of
right-handed neutrinos, etc.. In the original left-right symmetric
model \cite{ps1974} with doublet and bi-doublet Higgs scalars, the
right-handed Higgs doublet and the Higgs bi-doublet are responsible
for the left-right and electroweak symmetry breaking, respectively.
All of the fermions obtain Dirac masses through their Yukawa
couplings with the Higgs bi-doublet. In this sense it is difficult
to understand the small neutrino masses.

One possible solution is to replace the Higgs doublets by the Higgs
triplets \cite{ms1980}. The right-handed Higgs triplet develops a
large vacuum expectation value (VEV) for the heavy masses of the
left-handed Higgs triplet and the right-handed Majorana neutrinos.
The left-handed neutrinos then can obtain small Majorana masses
through the suppressed ratio of the electroweak scale over these
heavy masses. In this type-I
\cite{minkowski1977,yanadida1979,gps1979,glashow1980,ms1980} plus II
\cite{mw1980} seesaw context, the CP-violation and
out-of-equilibrium decays of the right-handed Majorana neutrinos
\cite{fy1986} and the left-handed Higgs triplet \cite{mz1992,ms1998}
can generate a lepton asymmetry through the self-energy
\cite{fps1995,pilaftsis1997,ms1998} and vertex corrections
\cite{fy1986,hs2004}. This lepton asymmetry is partially converted
to a baryon asymmetry through sphaleron \cite{krs1985} so that we
can understand the matter-antimatter asymmetry in the universe. This
is the so-called leptogenesis
\cite{fy1986,lpy1986,mz1992,fps1995,pilaftsis1997,ms1998,hs2004,pilaftsis1999,di2006}
mechanism.

Alternatively, one can revive the original left-right symmetric
model by introducing singlet fermions \cite{mohapatra1986}. These
singlet fermions can have Yukawa couplings with the Higgs and lepton
doublets. So the seesaw for the small neutrino masses is available
in the presence of the Majorana masses of the singlet fermions.
Specifically, the small Majorana masses of the singlet fermions will
induce the inverse \cite{mv1986} and linear \cite{barr2003} seesaw
while the large ones will give the double \cite{mohapatra1986} and
linear \cite{barr2003} seesaw. Depending on the size of the Majorana
masses of the singlet fermions, the leptogenesis can be realized in
two ways: (1) the right-handed neutrinos and the singlet fermions
form three pairs of quasi-degenerate Majorana neutrinos; (2) the
right-handed neutrinos obtain Majorana masses through the seesaw
contributions from the decoupled singlet fermions. We now begin to
demonstrate these possibilities in details.

\vspace{2mm}

\section{The Model}

For simplicity we do not write down the full Lagrangian. Instead, we
only give the part that is relevant for our discussions,
\begin{eqnarray}
\label{lagrangian} \mathcal{L}&\supset& -y\bar{\psi}_L^{}\phi
\psi_R^{}-\tilde{y}\bar{\psi}_L^{}\tilde{\phi} \psi_R^{}
-f_L^{}\bar{\psi}_L^{}\chi_L^{}\xi_R^{}-f_R^{}\bar{\psi}_R^{}\chi_R^{}\xi_R^{c}\nonumber\\
&&-\frac{1}{2}m_\xi^{}\bar{\xi}_R^{c}\xi_R^{}-\mu\chi_L^\dagger\phi\chi_R^{}-\tilde{\mu}\chi_L^\dagger\tilde{\phi}\chi_R^{}+\textrm{H.c.}\,.
\end{eqnarray}
Here $\psi_L^{}$ and $\psi_R^{}$ denote the left- and right-handed
lepton doublets for each family, $\chi_L^{}$ and $\chi_R^{}$ are the
left- and right-handed Higgs doublets, $\phi$ is the Higgs
bi-doublet, $\xi_R^{}$ stands for the three singlet fermions. We define
the left-right discrete symmetry to be the parity, under which the
fields transform as
\begin{subequations}
\label{parity}
\begin{eqnarray}
\psi_{L}^{}\equiv
\left(\textbf{2},\textbf{1},-1\right)&\stackrel{\mathcal{P}}{\longleftrightarrow}&
\psi_{R}^{}\equiv \left(\textbf{1},\textbf{2},-1\right)\,,\\
\chi_{L}^{}\equiv
\left(\textbf{2},\textbf{1},-1\right)&\stackrel{\mathcal{P}}{\longleftrightarrow}&
\chi_{R}^{}\equiv \left(\textbf{1},\textbf{2},-1\right)\,,\\
\phi
\equiv\left(\textbf{2},\textbf{2}^\ast_{},0\right)&\stackrel{\mathcal{P}}{\longleftrightarrow}&\phi^\dagger_{}
\equiv\left(\textbf{2}^\ast_{},\textbf{2},0\right)\,,\\
\xi_R^{}\equiv\left(\textbf{1},\textbf{1},0\right)&\stackrel{\mathcal{P}}{\longleftrightarrow}&
\xi_R^c\equiv\left(\textbf{1},\textbf{1},0\right)\,.
\end{eqnarray}
\end{subequations}
This constrains $y=y^\dagger_{}$, $\tilde{y}=\tilde{y}^\dagger_{}$,
$f_L^{}=f_R^{}=f$, $\mu=\mu^\ast_{}$ and
$\tilde{\mu}=\tilde{\mu}^\ast_{}$ in Eq. (\ref{lagrangian}).

After the right-handed Higgs doublet $\chi_R^{}$ develops its VEV
$\langle\chi_R^{}\rangle$, the $SU(2)_{L}^{}\times
SU(2)_{R}^{}\times U(1)_{B-L}^{}$ left-right symmetry is broken down
to the $SU(2)_{L}^{}\times U(1)_{Y}^{}$ electroweak symmetry. At
this stage, we can conveniently divide the Higgs bi-doublet into two
Higgs doublets: one will have a nonzero VEV for the electroweak
symmetry breaking and can be identified to the SM one, whereas the
other one will not develop any nonzero VEVs but can have a mass of
the order of the left-right breaking scale. For this purpose, it is
easy to describe
\begin{eqnarray}
\phi=[\phi_1^{},-\tilde{\phi}_2^{}]
\end{eqnarray}
and then define
\begin{eqnarray}
\varphi=\frac{\langle\phi_1^{}\rangle\phi_1^{}+\langle\phi_2^{}\rangle\phi_2^{}}{\sqrt{\langle\phi_1^{}\rangle^2_{}+\langle\phi_2^{}\rangle^2_{}}}\,,\quad
\eta
=\frac{\langle\phi_1^{}\rangle\phi_2^{}-\langle\phi_2^{}\rangle\phi_1^{}}{\sqrt{\langle\phi_1^{}\rangle^2_{}+\langle\phi_2^{}\rangle^2_{}}}
\end{eqnarray}
with
\begin{eqnarray}
\langle\varphi\rangle=\sqrt{\langle\phi_1^{}\rangle^2_{}+\langle\phi_2^{}\rangle^2_{}}\,,\quad
\langle\eta\rangle=0
\end{eqnarray}
for the electroweak symmetry breaking. Thus the right-handed
neutrinos $\nu_R^{}$ can have the Yukawa couplings to the SM Higgs
doublet $\varphi$, given by
\begin{eqnarray}
\label{yukawa1}
\mathcal{L}&\supset&-h_\nu^{}\bar{\psi}_L^{}\varphi
\nu_R^{}+\textrm{H.c.}
\end{eqnarray}
with
\begin{eqnarray}
h_\nu^{}=  y\frac{\langle\phi_1^{}\rangle}{\langle\varphi\rangle}
+\tilde{y}\frac{\langle\phi_2^{}\rangle} {\langle\varphi\rangle}\,.
\end{eqnarray}
On the other hand, the right-handed neutrinos $\nu_{R}^{}$ and the
singlet fermions $\xi_{R}^{}$ mix together as
\begin{eqnarray}
\label{massterm} \mathcal{L}\supset
-f\langle\chi_R^{}\rangle\bar{\nu}_R^{}\xi_R^{c}-\frac{1}{2}m_\xi^{}\bar{\xi}_R^{c}\xi_R^{}
+\textrm{H.c.}\,.
\end{eqnarray}
Clearly, the mass eigenstates, determined by a linear combination of
$\nu_{R}^{}$ and $\xi_{R}^{}$, can have very different properties
depending on the size of the Majorana mass term $m_\xi^{}$. For
example, $\nu_{R}^{}$ and $\xi_{R}^{}$ can form three pairs of
quasi-degenerate Majorana fermions if $f\langle\chi_R^{}\rangle\gg
m_\xi^{}$. On the contrary, with $f\langle\chi_R^{}\rangle\ll
m_\xi^{}$, we can give $\nu_{R}^{}$ Majorana masses due to the
seesaw contributions from $\xi_R^{}$. In the following we will study
the realization of leptogenesis and seesaw in the two limiting
cases.

\vspace{2mm}

\section{Leptogenesis with Linear and Inverse Seesaw}

In the case with $f\langle\chi_R^{}\rangle\gg m_\xi^{}$, we can
conveniently choose the base, where $f$ is rotated to be diagonal
and real, i.e.
$\displaystyle{f=\textrm{diag}\{f_1^{},f_2^{},f_3^{}\}}$, and then
diagonalize the mass terms (\ref{massterm}) by taking the rotations
as below,
\begin{subequations}
\begin{eqnarray}
\label{singlet1}
\nu_{R_i^{}}^{}&\simeq&\frac{1}{\sqrt{2}}\left(N_{R_i^{}}^{+}-iN_{R_i^{}}^{-}\right)\,,\\
\label{singlet2}
\xi_{R_i^{}}^{}&\simeq&\frac{1}{\sqrt{2}}\left(N_{R_i^{}}^{+}+iN_{R_i^{}}^{-}\right)\,.
\end{eqnarray}
\end{subequations}
Consequently there will be two physical Majorana fermions,
\begin{subequations}
\begin{eqnarray}
\label{majoranafermion1}
N_i^{+} &=& N_{R_i^{}}^{+}+ N_{R_i^{}}^{+c}\,,\\
\label{majoranafermion2} N_i^{-} &=& N_{R_i^{}}^{-}+ N_{R_i^{}}^{-c}
\end{eqnarray}
\end{subequations}
with
\begin{subequations}
\begin{eqnarray}
\label{mass1} m_{N^+_{i}}^{}&\simeq&f_{i}^{}v_\chi^{}
+ \frac{1}{2}m_{\xi_{ii}^{}}^{}\,,\\
\label{mass2} m_{N^-_{i}}^{}&\simeq &f_{i}^{}v_\chi^{} -
\frac{1}{2}m_{\xi_{ii}^{}}^{}\,.
\end{eqnarray}
\end{subequations}
The Yukawa couplings (\ref{yukawa1}) can be rewritten as
\begin{eqnarray}
\label{yukawa2} \mathcal{L}\supset-\frac{1}{\sqrt{2}}h_{\nu_{\alpha
i}^{}}^{}\bar{\psi}_{L_\alpha^{}}^{}\varphi N_i^{+} +
\frac{i}{\sqrt{2}}h_{\nu_{\alpha
i}^{}}^{}\bar{\psi}_{L_\alpha^{}}^{}\varphi N_i^{-} +
\textrm{H.c.}\,.
\end{eqnarray}

The singlet fermions $\xi_R^{}$ have Yukawa couplings to the
left-handed Higgs doublet $\chi_L^{}$ [the third term of Eq.
(\ref{lagrangian})]. Furthermore, $\chi_L^{}$ mixes with the SM
Higgs doublet $\varphi$,
\begin{eqnarray}
\label{mixing1} \mathcal{L}\supset-\frac{\left(\mu
\langle\phi_1^{}\rangle+\tilde{\mu}
\langle\phi_2^{}\rangle\right)\langle\chi_R^{}\rangle}{\langle\varphi\rangle}\chi^\dagger_{}\varphi
+ \textrm{H.c.}\,.
\end{eqnarray}
So, besides Eq. (\ref{yukawa2}), the Majorana fermions $N^{\pm}_{}$
can have other Yukawa couplings to $\varphi$ by integrating out
$\chi_L^{}$,
\begin{eqnarray}
\label{yukawa3} \mathcal{L}\supset-\frac{1}{\sqrt{2}}h_{\xi_{\alpha
i}^{}}^{}\bar{\psi}_{L_{\alpha}^{}}^{}\varphi N_i^+
-\frac{i}{\sqrt{2}}h_{\xi_{\alpha
i}^{}}^{}\bar{\psi}_{L_\alpha^{}}^{}\varphi N_i^- +\textrm{H.c.}
\end{eqnarray}
with
\begin{eqnarray}
\label{yukawa4} h_\xi^{}&=&-
f\frac{\langle\chi_R^{}\rangle\left(\mu\langle\phi_1^{}\rangle
+\tilde{\mu}\langle\phi_2^{}\rangle \right)}{m_{\chi_L^{}}^2
\langle\varphi\rangle}\,.
\end{eqnarray}
Here $m_{\chi_L^{}}^{}=\mathcal{O}(\langle\chi_R^{}\rangle)$ is the
mass of $\chi_L^{}$. Clearly, we have assumed $m_{\chi_L^{}}^{}\gg
m_{N_i^\pm}^{}$. This assumption is necessary for a successful
leptogenesis. For example, $\langle\chi_R^{}\rangle$ should be
bigger than $\mathcal{O}\left(10^{7}_{}\,\textrm{GeV}\right)$
\cite{mz1992,ky2004} to guarantee the departure from equilibrium of
$N_{1}^\pm$, if $N_{1}^\pm$ is the lightest pair with
$m_{N_{1}^\pm}^{}=\mathcal{O}\left(1-10\,\textrm{TeV}\right)$ to
account for the generation of a final lepton asymmetry. After the
electroweak symmetry breaking, $\varphi$ will develop a VEV
$\langle\varphi\rangle$ and then $\chi_L^{}$ will pick up a smaller
VEV,
\begin{eqnarray}
\langle\chi_L^{}\rangle\simeq-
\frac{\langle\chi_R^{}\rangle\left(\mu\langle\phi_1^{}\rangle
+\tilde{\mu}\langle\phi_2^{}\rangle \right)}{m_{\chi_L^{}}^2}\simeq
- \frac{\mu\langle\phi_1^{}\rangle
+\tilde{\mu}\langle\phi_2^{}\rangle }{\langle\chi_R^{}\rangle}\,.
\end{eqnarray}
The Yukawa couplings (\ref{yukawa4}) can simply be given by
\begin{eqnarray}
\label{yukawa5} h_\xi^{}&=&- f\frac{\langle\chi_L^{}\rangle}{
\langle\varphi\rangle}=-\textrm{diag}\{f_1^{},f_2^{},f_3^{}\}\frac{\langle\chi_L^{}\rangle}{
\langle\varphi\rangle}\nonumber\\
&=&\textrm{diag}\{h_{\xi_{e 1}^{}}^{},h_{\xi_{\mu
2}^{}}^{},h_{\xi_{\tau 3}^{}}^{}\}\,.
\end{eqnarray}

\subsection{Baryon Asymmetry}

Following the standard method \cite{pilaftsis1997} of the resonant
leptogenesis, we can compute the lepton asymmetry from the decays of
each $N_i^\pm$,
\begin{eqnarray}
\label{cpasymmetry} \varepsilon_{N_i^\pm}^{}
&=&\frac{\sum_{\alpha}^{}\left[\Gamma(N_i^\pm \rightarrow
\psi_{L_\alpha^{}}^{} + \varphi_{}^{\ast})-\Gamma(N_i^\pm
\rightarrow \psi_{L_\alpha^{}}^{c} +
\varphi)\right]}{\sum_{\alpha}^{}\left[\Gamma(N_1^\pm \rightarrow
\psi_{L_\alpha^{}}^{} + \varphi_{}^{\ast})+\Gamma(N_1^\pm
\rightarrow \psi_{L_\alpha^{}}^{c} +
\varphi)\right]}\nonumber\\
\vspace{10mm}
&\simeq&\frac{\left(h_\xi^{\dagger}h_\xi^{}-h_\nu^{\dagger}h_\nu^{}\right)_{ii}^{}
\textrm{Im}\left[\left(h_\nu^\dagger h_\xi^{}\right)_{ii}^{}\right]}
{4\pi A_{N_i^\pm}^{}}\frac{r_{N_i^{}}^{}}{r_{N_i^{}}^{2}
+\frac{1}{64\pi^2_{}}A_{N_i^\mp}^{2}}\nonumber\\
&&
\end{eqnarray}
with
\begin{eqnarray}
r_{N_i^{}}^{}&=&\frac{m_{N_i^{+}}^2-m_{N_i^{-}}^2}{m_{N_i^{+}}^{}m_{N_i^{-}}^{}}\simeq
\frac{2m_{\xi_{ii}^{}}}{f_i^{}\langle\chi_R^{}\rangle}\,,\\
A_{N_i^\pm}^{}&=&\frac{1}{2}\left[\left(h_\nu^{\dagger}\pm
h_\xi^{\dagger}\right)\left(h_\nu^{}\pm
h_\xi^{}\right)\right]_{ii}^{}\,.
\end{eqnarray}
From the calculation (\ref{cpasymmetry}), we emphasize that the
Yukawa couplings (\ref{yukawa2}) and (\ref{yukawa3}) are both
necessary \footnote{In the $SO(10)\rightarrow SU(5) \rightarrow
SU(2)_L^{}\times U(1)_Y^{}$ models for the resonant leptogenesis,
one could consider other possibilities \cite{ab2003} to make the
right-handed neutrinos and the singlet fermions both having the
Yukawa couplings to the SM lepton and Higgs doublets.} to generate
a nonzero lepton asymmetry. For demonstration, we assume $N_1^\pm$
to be much lighter than $N_{2,3}^\pm$. This means that the final
lepton and baryon asymmetry should mainly come from the decays of
$N_1^\pm$. In the weak and strong washout region, the final baryon
asymmetry can be approximately given by \cite{kt1990}
\begin{eqnarray}
\eta_{B}^{}&=&\frac{n_B^{}}{s}\nonumber\\
&=&-\frac{28}{79}\times\left\{\begin{array}{ll}\displaystyle{\frac{\varepsilon_{N_1^\pm}^{}}{g_\ast^{}}}&\textrm{for}~
K_{N_1^\pm}^{} \ll
1\,,\\
\\
\displaystyle{\frac{0.3\varepsilon_{N_1^+}^{}}{g_\ast^{}K_{N_1^\pm}^{}(\ln
K_{N_1^\pm}^{})^{0.6}_{}}}&\textrm{for}~ K_{N_1^\pm}^{} \gg 1
\end{array} \right.
\end{eqnarray}
with $g_{\ast}^{}\simeq 106.75$ being the relativistic degrees of
freedom (the SM fields). Here the quantity
\begin{eqnarray}
\label{ooe}
K_{N_1^\pm}^{}=\frac{\Gamma_{N_1^\pm}^{}}{2H(T)}\left|_{T=m_{N_1^\pm}^{}}^{}\right.
\end{eqnarray}
measures the effectiveness of the decays of $N_1^\pm$ at the
leptogenesis epoch. $\Gamma_{N_1^\pm}^{}$ and $H(T)$ are the decay
width and the Hubble constant, respectively, i.e.
\begin{eqnarray}
\Gamma_{N_1^\pm}^{}
&=&\frac{1}{8\pi}A_{N_1^\pm}^{}m_{N_1^\pm}^{}\,,\\
H(T)&=&\left(\frac{8\pi^{3}_{}g_{\ast}^{}}{90}\right)^{\frac{1}{2}}_{}
\frac{T^{2}_{}}{M_{\textrm{Pl}}^{}}\,.
\end{eqnarray}

\subsection{Neutrino Masses}

After the electroweak symmetry breaking, it is easy to read the
neutrino mass matrix by making use of the seesaw formula
\cite{minkowski1977,yanadida1979,gps1979,glashow1980,ms1980},
\begin{eqnarray}
\mathcal{L} &\supset& -\frac{1}{2}\bar{\nu}_L^{} m_{\nu}^{}\nu_L^c
 + \textrm{H.c.}
\end{eqnarray}
with
\begin{eqnarray}
m_\nu^{}\simeq
h_\nu^{}\frac{1}{f^\ast_{}}m_\xi^{}\frac{1}{f^\dagger_{}}h_\nu^T
\frac{\langle\varphi\rangle^2}{\langle\chi_R^{}\rangle^2_{}}-\left(h_\nu^{}+h_\nu^T\right)
\frac{\langle\varphi\rangle
\langle\chi_L^{}\rangle}{\langle\chi_R^{}\rangle}\,.
\end{eqnarray}
The second term is the linear seesaw \cite{barr2003}. As for the
first term, it is the inverse seesaw \cite{mv1986} for
$f\langle\chi_R^{}\rangle\gg m_\xi^{}$. So we denote
\begin{eqnarray}
m_\nu^{} =m_\nu^{\textrm{Inverse}}+m_\nu^{\textrm{Linear}}\,.
\end{eqnarray}

\subsection{Parameter Choice}

For a successful leptogensis and seesaw, we need to choose the
parameter space including two types of the Yukawa couplings
($h_\nu^{}$ and $f$), four VEVs ($\langle\chi_R^{}\rangle$,
$\langle\chi_L^{}\rangle$ and $\langle\phi_{1,2}^{}\rangle$) and two
cubic couplings ($\mu$ and $\tilde{\mu}$).

Firstly, we take
\begin{eqnarray}
\langle\chi_R^{}\rangle=3\times 10^{7}_{}\,\textrm{GeV}
\end{eqnarray}
and then
\begin{subequations}
\label{parameter-mass}
\begin{eqnarray}
m_{N_{1}^\pm}^{}&=&~\,3\,\textrm{TeV}\quad
\textrm{for}\quad ~\,f_1^{}=10^{-4}_{}\,,\\
m_{N_{2,3}^\pm}^{}&=&30\,\textrm{TeV}\quad \textrm{for}\quad
f_{2,3}^{}=10^{-3}_{}\,.
\end{eqnarray}
\end{subequations}
This choice can guarantee that the gauge interactions of $N_{1}^\pm$
have been decoupled at the leptogenesis epoch $T\sim
m_{N_{1}^\pm}^{}$ \cite{mz1992,ky2004}. Secondly, we assume
\begin{eqnarray}
\mu\simeq \tilde{\mu}=\mathcal{O}(0.1)\langle\chi_R^{}\rangle
\end{eqnarray}
and then perform
\begin{eqnarray}
\label{parameter-vev}
\langle\chi_L^{}\rangle=0.1\langle\varphi\rangle\,.
\end{eqnarray}
From the above ratio, it is easy to solve
\begin{eqnarray}
\langle\varphi\rangle\simeq173\,\textrm{GeV}\,,\quad\langle\chi_L^{}\rangle\simeq17\,\textrm{GeV}
\end{eqnarray}
for
\begin{eqnarray}
\sqrt{\langle\varphi\rangle^2_{}+\langle\chi_L^{}\rangle^2_{}}\simeq
174\,\textrm{GeV}\,.
\end{eqnarray}
With the inputs (\ref{parameter-mass}) and (\ref{parameter-vev}),
Eq. (\ref{yukawa5}) induces
\begin{eqnarray}
h_{\xi_{e 1}^{}}^{}= 0.1\,h_{\xi_{\mu 2}^{}}^{}=0.1\,h_{\xi_{\tau
3}^{}}^{}=-10^{-5}_{}\,.
\end{eqnarray}
Thirdly, we consider
\begin{eqnarray}
m_{\xi_{ij}^{}}^{}=\mathcal{O}(0.1-1\,\textrm{keV})
\end{eqnarray}
to determine
\begin{eqnarray}
r_{N_1^{}}^{}&=&10^{-10}_{}\,.
\end{eqnarray}

Finally, we assume the linear seesaw dominates the neutrino mass
matrix. We further consider the simple case that the Yukawa
couplings $h_\nu^{}$ are symmetric, i.e. $h_\nu^{}=h_\nu^T$. In this
case, it is easy to determine $h_\nu^{}$ by
\begin{eqnarray}
h_\nu^{}=-\frac{\langle\chi_R^{}\rangle}{2\langle\varphi\rangle\langle\chi_L^{}\rangle}U^\ast_{\textrm{PMNS}}
\textrm{diag}\{m_1^{}\,,~m_2^{}\,,~m_3^{}\}
U^\dagger_{\textrm{PMNS}}\,.
\end{eqnarray}
Here $U_{\textrm{PMNS}}^{}$ is the Pontecorvo-Maki-Nakagawa-Sakata
\cite{mns1962} (PMNS) leptonic mixing matrix while $m_{1,2,3}^{}$
are the eigenvalues of the neutrino mass matrix. Currently the
neutrino oscillation data have precisely measured the two neutrino mass
squared differences \cite{stv2008},
\begin{eqnarray}
~\Delta m_{21}^2~ &=& ~m_2^2-m_1^2~=7.65^{+0.23}_{-0.20} \times
10^{-5}_{}\,\textrm{eV}^2_{}\,,\nonumber\\
\vspace{10mm} |\Delta m_{31}^2| &=&
|m_3^2-m_1^2|\,=2.40^{+0.12}_{-0.11}\times
10^{-3}_{}\,\textrm{eV}^2_{}\,.
\end{eqnarray}
As for the PMNS matrix, it is consistent with the tri-bimaximal
mixing \cite{hps2002},
\begin{widetext}
\begin{eqnarray}
\label{pmns} U_{\textrm{PMNS}}^{} =\left(\begin{array}{ccc}
~~\sqrt{\frac{2}{3}} & ~~\sqrt{\frac{1}{3}}
& 0 \\
[-2mm]~&~&~\\ -\sqrt{\frac{1}{6}} &
~~\sqrt{\frac{1}{3}} & \sqrt{\frac{1}{2}}\\
[-2mm]~&~&~\\ ~~\sqrt{\frac{1}{6}} & -\sqrt{\frac{1}{3}} &
\sqrt{\frac{1}{2}}
\end{array}\right)
\textrm{diag}\{e^{i\frac{\alpha_1^{}}{2}}_{}\,,~e^{i\frac{\alpha_2^{}}{2}}_{}\,,~1\}\,,
\end{eqnarray}
\end{widetext}
with which we can derive
\begin{widetext}
\begin{eqnarray}
h_\nu^{}=-\frac{\langle\chi_R^{}\rangle}{2\langle\varphi\rangle\langle\chi_L^{}\rangle}\left(\begin{array}{lll}
~~\,\frac{2}{3}\tilde{m}_1^{}+\frac{1}{3}\tilde{m}_2^{} &~
-\frac{1}{3}\tilde{m}_1^{}+\frac{2}{3}\tilde{m}_2^{}
& ~~\, ~~\frac{1}{3}\tilde{m}_1^{}-\frac{1}{3}\tilde{m}_2^{} \\
[-2mm]~&~&~\\ -\frac{1}{3}\tilde{m}_1^{}+\frac{2}{3}\tilde{m}_2^{}
&~
-\frac{1}{6}\tilde{m}_1^{}+\frac{1}{3}\tilde{m}_2^{}+\frac{1}{3}\tilde{m}_3^{} & ~-\frac{1}{6}\tilde{m}_1^{}-\frac{1}{3}\tilde{m}_2^{}+\frac{1}{3}\tilde{m}_3^{}\\
[-2mm]~&~&~\\~~\,\frac{1}{3}\tilde{m}_1^{}-\frac{1}{3}\tilde{m}_2^{}
&
~-\frac{1}{6}\tilde{m}_1^{}-\frac{1}{3}\tilde{m}_2^{}+\frac{1}{3}\tilde{m}_3^{}
&
~~~~\,\frac{1}{6}\tilde{m}_1^{}+\frac{1}{3}\tilde{m}_2^{}+\frac{1}{3}\tilde{m}_3^{}
\end{array}\right)~~~\textrm{with}~~~\left\{\begin{array}{l}
\tilde{m}_1^{}=m_1^{}e^{-i\alpha_1^{}}_{}\,,\\
[-2mm]~\\ \tilde{m}_2^{}=m_2^{}e^{-i\alpha_2^{}}_{} \,,\\
[-2mm]~\\ \tilde{m}_3^{}=m_3^{}\,.
\end{array}\right.
\end{eqnarray}
\end{widetext}
For demonstration, let's focus on the normal hierarchical neutrinos,
i.e.
\begin{eqnarray}
m_1^{}=0\,,~~m_2^{}=\sqrt{\Delta m_{21}^2}\,,~~m_3^{}=\sqrt{|\Delta
m_{31}^2|}\,.
\end{eqnarray}

It is easy to check that the linear seesaw dominates the neutrino mass
matrix with the above parameter choice. We also can fix the
parameters (\ref{ooe}) to be
\begin{eqnarray}
K_{N_1^+}^{}\simeq K_{N_1^-}^{}=242\,.
\end{eqnarray}
By further inputting the CP phase
\begin{eqnarray}
\sin\frac{h_{\nu_{e 1}^{}}^\ast h_{\xi_{e 1}^{}}^{}}{|h_{\nu_{e
1}^{}}^{} h_{\xi_{e 1}^{}}^{}|}=\sin\alpha_2^{}=-0.13\,,
\end{eqnarray}
the CP asymmetries (\ref{cpasymmetry}) can also be determined,
\begin{eqnarray}
\varepsilon_{N_1^+}^{}\simeq \varepsilon_{N_1^-}^{}=-2.99\times
10^{-5}_{}\,.
\end{eqnarray}
In consequence, the final baryon asymmetry should arrive at
\begin{eqnarray}
\eta_{B}^{}=0.886\times 10^{-10}_{}\,,
\end{eqnarray}
which is consistent with the five-year observations of the WMAP
collaboration \cite{dunkley2008},
\begin{eqnarray}
\label{cmb}
\eta_{B}^{}&=&\frac{1}{7.04}\times (6.225\pm0.170)\times 10^{-10}_{}\nonumber\\
&=&(0.884\pm0.024)\times 10^{-10}_{}\,.
\end{eqnarray}

\vspace{2mm}

\section{Leptogenesis with Double Seesaw}

We now discuss the case with $f\langle\chi_R^{}\rangle\gg m_\xi^{}$.
In this case, the mass terms (\ref{massterm}) can be diagonalized
into two blocks, i.e.
\begin{eqnarray}
\mathcal{L}\supset-\frac{1}{2}m_{N}^{}\bar{\nu}_R^c
\nu_R^{}-\frac{1}{2}m_\xi^{}\bar{\xi}_R^c\xi_R^{}+ \textrm{H.c.}\,.
\end{eqnarray}
Here the Majorana mass matrix of the right-handed neutrinos is a
seesaw solution,
\begin{eqnarray}
m_N^{}=-f^\ast_{}\frac{\langle\chi_R^{}\rangle^2_{}}{m_\xi^\dagger}f^\dagger_{}\,,
\end{eqnarray}
which can be of the order of
$m_N^{}=\mathcal{O}(10^{3}_{}-10^{10}_{}\,\textrm{GeV})$ for
$m_\xi^{}=\mathcal{O}(10^{16}_{}-10^{19}_{}\,\textrm{GeV})$ and
$f\langle\chi_R^{}\rangle=\mathcal{O}(10^{12}_{}-10^{13}_{}\,\textrm{GeV})$.
The Yukawa interaction (\ref{yukawa1}) can thus realize the
leptogenesis, with or without resonant effect, in the traditional way.
The right-handed neutrinos are much lighter than the right-handed
gauge bosons so that their gauge interactions can be
decoupled naturally for the out-of-equilibrium condition.

The right-handed neutrinos and the singlet fermions can both
contribute to the neutrino masses,
\begin{eqnarray}
\mathcal{L} &\supset& -\frac{1}{2}\bar{\nu}_L^{} m_{\nu}^{}\nu_L^c
 + \textrm{H.c.}\,.
\end{eqnarray}
with
\begin{eqnarray}
m_\nu^{}&\simeq &
h_\nu^{}\frac{1}{f^\ast_{}}m_\xi^{}\frac{1}{f^\dagger_{}}h_\nu^T
\frac{\langle\varphi\rangle^2}{\langle\chi_R^{}\rangle^2_{}}-\left(h_\nu^{}+h_\nu^T\right)
\frac{\langle\varphi\rangle
\langle\chi_L^{}\rangle}{\langle\chi_R^{}\rangle}\nonumber\\
&=&-h_\nu^{}\frac{\langle\varphi\rangle^2_{}}{m_N^{}}h_\nu^T-\left(h_\nu^{}+h_\nu^T\right)
\frac{\langle\varphi\rangle
\langle\chi_L^{}\rangle}{\langle\chi_R^{}\rangle}\,.
\end{eqnarray}
The first term is usually called the double seesaw
\cite{mohapatra1986} since there are two seesaw steps generating the
small neutrino masses. The neutrino mass term can be conveniently
expressed by
\begin{eqnarray}
m_\nu^{} =m_\nu^{\textrm{Double}}+m_\nu^{\textrm{Linear}}\,.
\end{eqnarray}
The linear seesaw $m_\nu^{\textrm{Linear}}$ could be comparable to
the double seesaw $m_\nu^{\textrm{Double}}$. This implies we could
relax the constraint on the CP asymmetry in the decays of the
right-handed neutrinos from the neutrino masses if there is a
cancellation between $m_\nu^{\textrm{Double}}$ and
$m_\nu^{\textrm{Linear}}$.

\vspace{10mm}

\section{Summary}

\vspace{-2mm}

In summary we have discussed the realization of leptogenesis and
seesaw by adding three singlet fermions with Majorana masses in the
original left-right symmetric model with doublet and bi-doublet
Higgs scalars. Depending on the size of the Majorana masses of the
singlet fermions, the leptogenesis can be realized in different
seesaw scenarios. If the Majorana masses are small, the right-handed
neutrinos and the singlet fermions can form three pairs of
quasi-degenerate Majorana fermions to accommodate the
resonant leptogenesis at the TeV scale naturally. The neutrino masses are
dominated by the linear seesaw, although the inverse seesaw also
exists. In the other limiting case, where the singlet fermions are
very heavy, the right-handed neutrinos can obtain Majorana masses
through the seesaw contribution from the singlet fermions and then
give a traditional picture of leptogenesis.

\vspace{5mm}

\textbf{Acknowledgement}: PHG thanks Manfred Lindner for hospitality
at Max-Planck-Institut f\"{u}r Kernphysik and thanks the Alexander
von Humboldt Foundation for financial support.

\end{document}